\documentclass[sigconf]{acmart}

\usepackage{colortbl}

\AtBeginDocument{%
  \providecommand\BibTeX{{%
    \normalfont B\kern-0.5em{\scshape i\kern-0.25em b}\kern-0.8em\TeX}}}

\copyrightyear{2023} 
\acmYear{2023} 
\setcopyright{rightsretained} 
\acmConference[WWW '23]{Proceedings of the ACM Web Conference 2023}{April 30-May 4, 2023}{Austin, TX, USA}
\acmBooktitle{Proceedings of the ACM Web Conference 2023 (WWW '23), April 30-May 4, 2023, Austin, TX, USA}
\acmDOI{10.1145/3543507.3583388}
\acmISBN{978-1-4503-9416-1/23/04}


\begin{document}

\title{Reinforcement Learning-based Counter-Misinformation Response Generation: A Case Study of COVID-19 Vaccine Misinformation}

\author{Bing He, Mustaque Ahamad, Srijan Kumar}
\affiliation{%
  \institution{Georgia Institute of Technology}
  \city{Atlanta}
  \state{Georgia}
  \country{USA}
}
\email{bhe46@gatech.edu, mustaq@cc.gatech.edu, srijan@gatech.edu}



\begin{abstract}
The spread of online misinformation threatens public health, democracy, and the broader society. 
While professional fact-checkers form the first line of defense by fact-checking popular false claims, they do not engage directly in conversations with misinformation spreaders.
On the other hand, non-expert ordinary users act as eyes-on-the-ground who proactively counter misinformation -- recent research has shown that 96\% counter-misinformation responses are made by ordinary users. 
However, research also found that 2/3 times, these responses are rude and lack evidence. 
This work seeks to create a counter-misinformation response generation model to empower users to effectively correct misinformation. 
This objective is challenging due to the absence of datasets containing ground-truth of ideal
counter-misinformation responses, and the lack of models that can generate responses backed by communication theories. 
In this work, we create two novel datasets of misinformation and counter-misinformation response pairs from in-the-wild social media and crowdsourcing from college-educated students. 
We annotate the collected data to distinguish poor from ideal responses that are factual, polite, and refute misinformation.
We propose MisinfoCorrect, a reinforcement learning-based framework that learns to generate counter-misinformation responses for an input misinformation post. 
The model rewards the generator to increase the politeness, factuality, and refutation attitude while retaining text fluency and relevancy. 
Quantitative and qualitative evaluation shows that our model outperforms several baselines by generating high-quality counter-responses. This work illustrates the promise of generative text models for social good -- here, to help create a safe and reliable information ecosystem. The code and data is accessible on \url{https://github.com/claws-lab/MisinfoCorrect}.

\end{abstract}


\settopmatter{printfolios=true} 
\fancyhead{} 

\maketitle

\begin{figure}[!t]
    \centering
    \includegraphics[width=0.48\textwidth]{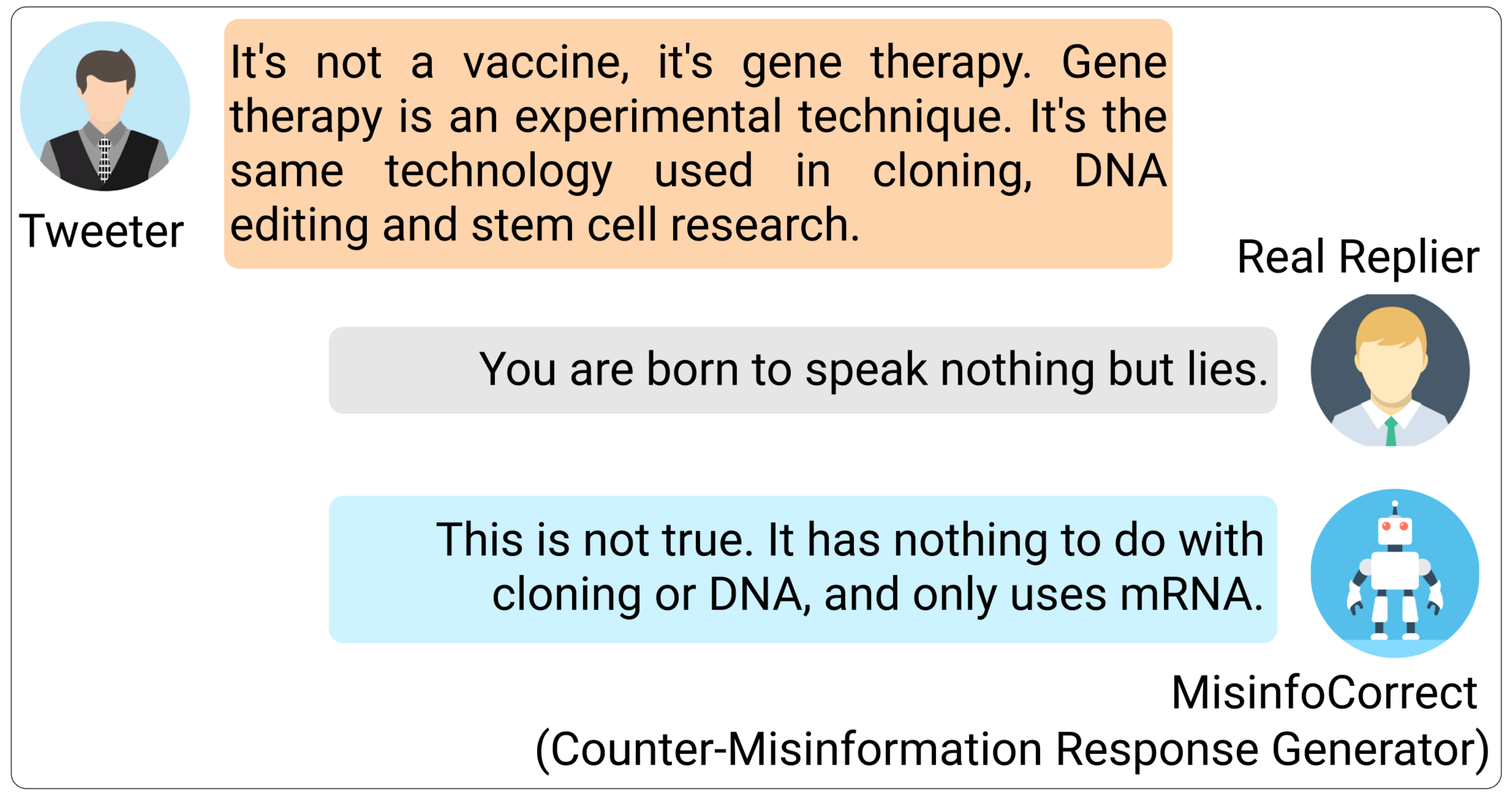}
    \vspace{-7mm}
    \caption{An overview of counter-misinformation response generation task. 
    } 
    \vspace{-4mm}
    \label{fig:task_demo}
\end{figure}

\section{Introduction}

Online misinformation reduces trust in vaccines and health policies~\cite{pierri2022online,ball2020epic,lazer2018science}, leads to violence and harassment~\cite{starbird2014rumors,arif2018acting}, questions democratic processes and elections~\cite{silverman2015lies,silverman2016analysis,shin2017partisan}, increases polarization~\cite{stewart2018examining}, and harms well-being~\cite{verma2022examining}.
Most people receive information and news from social media~\cite{walker2021news}, which is often ``ground-zero'' for health misinformation and where misinformation spreads faster and farther than truth~\cite{vosoughi2018spread,lazer2018science}. 
COVID-19 vaccine misinformation, including false claims that the vaccine causes infertility, contains microchips and even changes DNA and genes has fueled vaccine hesitancy, reduced vaccine uptake, and prolonged the pandemic. Besides, misinformation also causes harms to people directly. For example, misinformation that Bill Gates creates vaccines to depopulate people led to distrust and verbal attacks~\cite{fuchs2021bill}.
Thus, it is critical to curb the spread of online misinformation~\cite{Micallef2020, lewandowsky2012misinformation, Goindani2020, Budak2011, Zhu2021, Litou2017, Wang2020}. 
In this work, we use a broad definition of misinformation which includes falsehoods, inaccuracies, rumors, decontextualized truths, or misleading leaps of logic~\cite{kumar2018false,wu2019misinformation}.

Professional fact-checkers and journalists provide objective fact-checks for viral claims and release their determination on their website, which are incredibly useful to create detection models. However, fact-checkers do not actively engage with misinformation spreaders on social media platforms~\cite{Micallef2020}. 
On the other hand, non-expert social media users, i.e., ordinary users or crowd, act as eyes-on-the-ground who proactively question and counter misinformation, including emerging misinformation~\cite{Micallef2020,Tully2020,vo2018rise,zhou2022fake,Bode2018,starbird2014rumors,  miyazaki2022fake}. They complement fact checkers who can only verify a handful of stories after they have gone viral~\cite{Allen2021,kim2018leveraging}.
Recent evidence shows that 96\% of counter-misinformation responses are made by ordinary users, while professionals account for the rest~\cite{Micallef2020}.

Recent research on social correction~\cite{ma2023characterizing}, i.e., countering  of misinformation claims by other social media users, has proven to be as effective as professional correction~\cite{seo2022if}, 
curbs misinformation spread~\cite{friggeri2014rumor,colliander2019fake,Wijenayake2021},
and works across topics~\cite{bode2020right,vraga2018not,vraga2021addressing,bode2015in,Bode2018,vraga2020correction}, platforms and demographics~\cite{vraga2021addressing,vraga2021assessing,vraga2020testing,vraga2021effects}.
Corrections work~\cite{chan2017debunking,walter2021evaluating,walter2020fact,walter2018unring,porter2021global} without causing an increase in misperception (i.e., the backfire effect has not been replicated)~\cite{swire2020searching,guess2020does,wood2019elusive}.
While corrections are not expected to convince everyone, 
they are most effective in reducing misinformation consumers' misperceptions~\cite{Bode2021,colliander2019fake,Bode2018,Wijenayake2021,seo2021effectiveness}.
Thus, empowering users to effectively correct misinformation promises a scalable solution towards information integrity. 
This solution is independent of, but complements, the efforts of social media platforms to detect misinformation via the crowd, e.g., Twitter Birdwatch~\cite{prollochs2022community}.

Alarmingly, linguistic analyses of in-the-wild crowd-generated counter-responses revealed that 2 out of 3 counter-misinformation posts are rude and do not use fact-checking evidence to support their counter-response~\cite{Micallef2020}. Uncivil counter-responses can lead to reduced trust in the correcting user~\cite{flekova2016exploring,thorson2010credibility} and result in arguments~\cite{Masullo2021,kumar2018community,cheng2017anyone}. 
This implies an urgent need to empower crowds so that they counter misinformation more effectively.

Thus, in this work, we seek to facilitate healthy misinformation correction by the crowd, which includes being objective, evidenced, and polite -- properties that have been shown to be effective~\cite{Tanaka2019,seo2022if}.
To do so, we propose to create a counter-misinformation response generator, which generates desirable counter-response for a misinformation post (as illustrated in Figure~\ref{fig:task_demo}). 
Our study is focused on countering misinformation on Twitter, given its prominence in the spread of online misinformation.

\textbf{Challenges.} Generating effective counter-misinformation responses poses several challenges. 
\underline{First}, there is no existing dataset containing pairs of annotated misinformation posts and counter responses. 
\underline{Second}, there is no counter-misinformation response generator model. The closest research works in fact-checking generator~\cite{vo2019learning} are non-conversational and related research in counter-hate speech/counter-argument generator~\cite{tekiroglu2020generating, zhu2021generate, chung2021towards, alshomary2021counter} do not apply directly since they are not evidence-based or not specific to misinformation.
\underline{Third}, counter-misinformation responses are effective if they have the following desirable properties: 
objective and evidenced~\cite{Tanaka2019,seo2022if}, 
makes rational arguments~\cite{Orosz2016}, 
refutes fallacy in reasoning~\cite{Stojanov2015}, 
and polite~\cite{steffens2019organisations, malhotra2022meaning}. 
Off-the-shelf text generator models do not directly generate counter-responses with this desiderata. 
\underline{Four}, bot-generated or template-based responses are not effective since they are non-personalized and non-contextualized with respect to the false claims made in the misinformation post. Thus, the counter-response needs to be relevant to the misinformation post.

\textbf{Present work.} 
We propose to create two novel datasets containing misinformation and counter-responses (solution to challenge 1) -- one collected from in-the-wild social media responses from Twitter and another created by crowd-sourcing from college students.
We focus on four popular COVID-19 vaccine misinformation topics on Twitter (e.g., Bill Gates created vaccines to depopulate people~\cite{sharma2022covid, evanega2020coronavirus}, and vaccines can cause infertility~\cite{abbasi2022widespread, hsu2022sources}, contain microchip~\cite{skafle2022misinformation}, alter DNA~\cite{loomba2021measuring, nuzhath2020covid}).
To create the in-the-wild dataset, for each misinformation topic, we collect all the replies to misinformation tweets identified in prior research~\cite{hayawi2022anti}. We annotate associated replies to identify the responses that counter the tweet along with their textual attributes of refuting, politeness, and factuality. Finally, we have 754 misinformation tweet and countering response pairs.
For the crowd-sourced response generation, we recruit and train 17 college students to write counter-misinformation replies when given misinformation posts. 
In total, we collect 591 crowdsourced replies.

Next, we propose a reinforcement learning-based framework, called MisinfoCorrect, that learns to generate counter-misinformation responses that are polite, evidenced, and refute misinformation (solutions to challenges 2 and 3). 
Specifically, this agent utilizes a policy network on a transformer-based language model adapted from GPT-2~\cite{radford2019language}. 
During training, we reward the generation that increases the politeness and refutation attitude.
Additionally, we ensure text fluency and relevancy to the misinformation post by adding fluency and relevance rewards in the reinforcement learning framework (solution to challenge 4). 

MisinfoCorrect is evaluated against five representative baselines on the task of counter-misinformation response generation. 
Quantitative and qualitative experiments show that it can outperform the baselines by generating high-quality counter-responses. 

To summarize, our contributions are as follows:

\noindent $\bullet$ We create two large novel and annotated datasets containing misinformation and counter-response pairs from social media (in-the-wild) and generated via crowd-sourcing (in-lab). Together, both datasets contain 1,345 counter-misinformation responses. 

\noindent $\bullet$ We propose a reinforcement learning based counter-response generation framework, where the counter-response is especially rewarded for being polite, evidenced, and refuting misinformation. 

\noindent $\bullet$ Results on actual COVID-19 vaccine misinformation conversations show that the proposed model outperforms existing representative baselines. 

The code and data is accessible on \url{https://github.com/claws-lab/MisinfoCorrect}.
\section{Related Works}

\subsection{Social Correction of Misinformation by Non-Expert Ordinary Users}

Recent studies have shown remarkable effectiveness of social correction by non-expert users by conducting experiments via interviews~\cite{Borah2021,Kirchner2020,Tully2020}, surveys~\cite{Veeriah2021,Kirchner2020}, and in-lab experiments~\cite{Tully2020}. 
This correction has been shown to be as effective as professional correction~\cite{seo2022if}, 
curbs misinformation spread~\cite{friggeri2014rumor,colliander2019fake,Wijenayake2021},
and works across topics~\cite{bode2020right,vraga2018not,vraga2021addressing,bode2015in,Bode2018,vraga2020correction}, platforms and demographics~\cite{vraga2021addressing,vraga2021assessing,vraga2020testing,vraga2021effects}.
Notably, users' polite and evidenced responses that refute misinformation are shown to effectively counter misinformation and reduce the belief in misinformation~\cite{steffens2019organisations, malhotra2022meaning, Tanaka2019, seo2022if, Orosz2016, Orosz2016, Stojanov2015, chan2017debunking}.
Users correct others, typically friends~\cite{margolin2018political}, owing to a sense of social duty~\cite{friggeri2014rumor, Veeriah2021, Grandhi2021, mosleh2021perverse, Pal2019}, anger, or guilt ~\cite{Sun2021}.
These works provide considerable evidence that correction by ordinary users is effective when countering misinformation and in mitigating the spread of misinformation. On the other hand, considering the limited capability of professional fact-checkers, the large number of ordinary users and their efforts in social correction show great potential for a scalable solution to countering misinformation.

\subsection{Analysis of Crowd-Generated Misinformation Flagging and Countering}
Crowd-generated counter-misinformation complements fact-checking and correction by professionals -- the latter has already been studied extensively~\cite{seo2022if, hameleers2020misinformation, hameleers2022separating, zhang2021effects, walter2020fact, pavleska2018performance}.
Emerging research has analyzed the role that non-experts play in flagging and countering misinformation. 
Twitter's Birdwatch~\cite{prollochs2022community} is a recently-launched platform that allows users to report and flag misinformation. Studies have analyzed the data from Twitter Birdwatch~\cite{prollochs2022community,miyazaki2022fake,drolsbach2022diffusion,allen2022birds}, which have shown how users actively engage to identify tweets that they believe are misleading and provide contextual notes to debunk them. Users have different levels of debunking capability.
However, Birdwatch only allows users to flag misinformation and does not allow user-to-user communication and countering of misinformation on Twitter. 
Thus, user flagging behavior within the Birdwatch ecosystem is not representative of user behavior on the broader Twitter platform or on other social media platforms. 
Recent works by \citet{Micallef2020,micallef2022cross} have analyzed how users counter misinformation in-the-wild on Twitter, Facebook, and Reddit. They showed that 96\% of all counter misinformation posts on Twitter are made by ``ordinary citizens''~\cite{Micallef2020} and counter-misinformation behavior happens on multiple platforms~\cite{micallef2022cross}. 
Existing works, however, have not studied how to empower the crowd to counter and correct misinformation by generating effective responses.

\subsection{Fact-Check Generation Methods}
The goal of fact-check generation methods~\cite{vo2019learning,vo2020standing} is to respond to misinformation with a fact-checking URL.
However, we consider a broader task of counter-response generation where the response text has to be generated. 
Existing works~\cite{vo2019learning,vo2020standing} consider any post with a fact-checking URL to two websites (Snopes and Politifact) as a fact-checking response, which is an inaccurate assumption -- a fact-checking URL can be present to ridicule or oppose the fact-check~\cite{micallef2022cross} and can be taken out of context~\cite{micallef2022cross}. 
Importantly, only 1 out of 3 users use URL evidence when correcting misinformation~\cite{Micallef2020} and YouTube is the most frequently used URL, instead of fact-checking URLs~\cite{micallef2022cross}; consequently, studies relying only on fact-checking URLs are limited in their scope and do not learn from the majority of user-generated corrective posts. 
Our work overcomes these shortcomings by creating two novel datasets (Section~\ref{sec:data}), one using social media, including both URL and non-URL responses, and another using crowdsourced data collection. We further perform several manual annotation steps while creating the data to ensure only exact counter-responses are present in the data.

\subsection{Counter-Hate and Counter-Argument Text Generation}
Counter-hate~\cite{tekiroglu2020generating, zhu2021generate, chung2021towards, he2021racism} and counter-argument~\cite{alshomary2021counter, hua2019argument, hidey2019fixed} text generation tasks are also related to our problem setting, where the generated text is aimed to refute the original post spreading hate and any generic argument, respectively. Some proposed models fine-tune large scale unsupervised language models on the hate-speech or argument text for text generation~\cite{tekiroglu2020generating, schiller2020aspect}. Other models first generate a set of candidate counter-hate/counter-argument replies, and then select one based on the relevance to the original post in a generate-then-retrieve or identify-substitute manner~\cite{zhu2021generate, hua2019argument, hidey2019fixed} . Meanwhile, some related counter-hate/counter-argument datasets have also been  released~\cite{qian2019benchmark, saakyan2021covid, hua2019argument}. 
However, it should be noted that compared to counter-misinformation response generation, the task of counter-hate generation does not necessitate responses to be evidence-based. Similarly, the counter-argument generation is a generic task (e.g., arguing whether immigration is good) and is not specific to misinformation.
Additionally, large annotated and curated datasets exist for counter-hate and counter-argument~\cite{qian2019benchmark, saakyan2021covid}, which is not the case for counter-misinformation generation. To fill these gaps, we both curate two novel datasets and propose a counter-misinformation generator which can refute misinformation while being polite and providing evidence.

\section{Problem Definition}\label{sec:prob_definition}

Given a misinformation post $m$, we aim to build a text generator $g$ such that it can output counter-response $\hat{c} = g(m)$, which has certain desirable properties $\mathcal{P}$.

The \textbf{desirable properties} of $\hat{c}$ are motivated by research works from social scientists, journalists and psychologists regarding misinformation correction, which shows that counter responses are effective if they have the following desirable properties: politeness~\cite{steffens2019organisations, malhotra2022meaning}, objective and evidenced~\cite{Tanaka2019,seo2022if}, make rational arguments~\cite{Orosz2016}, convey the competence of the commenter~\cite{Orosz2016}, and refute fallacy in reasoning~\cite{Stojanov2015, chan2017debunking}.
More elaborately, the desirable properties include: 
\begin{itemize}
    \item \textit{Refuting:} the response explicitly refutes the the misinformation to correct the misinformation spreader. The expressed refutation via explicitly and objectively refuting misinformation in counter response can reduce misinformation's impact~\cite{Tanaka2019}.
    \item \textit{Evidence:} the response contains supporting sentences to back up the refutation. Evidenced-based responses can more effectively debunk the misleading claims, and likely reduce the belief of misinformation poster~\cite{chan2017debunking}. More importantly, people are more willing to agree with a countering response when it is evidence-based~\cite{chan2017debunking}. 
    \item \textit{Politeness:} the response is polite to avoid possible backfire. When countering misinformation, uncivil responses can aggravate the misinformation poster, while it is  more likely that the misinformation spreader favorably considers the true information when responses are polite~\cite{steffens2019organisations, malhotra2022meaning}. 
\end{itemize}
Beyond these specific requirement in misinformation correction domain, other textual properties are also required in generated text:
\begin{itemize}
    \item \textit{Fluency:} the generated text should be fluent in expression such that it is natural for people to read and understand.
    \item \textit{Relevance:} the response should be relevant to the misinformation post and ensure coherent expression.
\end{itemize}

\section{Counter-Response Datasets: In-the-Wild and Crowdsourced}
\label{sec:data}
We create two novel counter-response datasets, first containing in-the-wild social media counter-responses and second containing crowdsourced in-lab counter-responses.

\subsection{Misinformation Topics}
\label{sec:misinfotopics}

We focus on COVID-19 vaccine misinformation due to its impact across the world. We mainly choose four popular misinformation topics to which a large number of users have been exposed and impacted~\cite{Micallef2020, sharma2022covid, evanega2020coronavirus, skafle2022misinformation, abbasi2022widespread, hsu2022sources, loomba2021measuring, nuzhath2020covid}. These misinformation topics gained popularity from December 2020 when the COVID-19 vaccines were approved by the FDA~\cite{sharma2022covid}, in essence, 
\noindent $\bullet$ Bill Gates conspiracy theories~\cite{sharma2022covid, evanega2020coronavirus}: This includes conspiracies claiming that Bill Gates created the COVID-19 vaccine to depopulate people or he holds the patents for COVID-19 vaccine to profit from the vaccine sales. 

\noindent $\bullet$ COVID-19 vaccines contain microchips to track people~\cite{skafle2022misinformation}. 

\noindent $\bullet$ COVID-19 vaccines cause infertility and prevent pregnancy in women~\cite{abbasi2022widespread, hsu2022sources}. 

\noindent $\bullet$ COVID-19 vaccines alter DNA or the vaccine is gene therapy~\cite{loomba2021measuring}. 

\subsection{In-the-wild Social Media Counter-Response Dataset}\label{data:annotated_data_in_the_wild_all}

\subsubsection{\textbf{Misinformation Tweet and Response Collection}}\label{sec:misinfo_data_in_the_wild}

Our dataset builds on $14,123,473$ COVID-19 vaccine-related tweets crawled by~\citet{hayawi2022anti} from Dec 1, 2020 to July 31, 2021. Since we are more focused on responses rather than tweets themselves, we only keep tweets having at least one response, resulting in $1,609,069$ tweets. 

To identify misinformation tweets, we first create a COVID-19 vaccine misinformation tweet classifier using BERT~\cite{devlin2018bert} based on tweet annotations provided by~\citet{hayawi2022anti}. This classifier has a performance in precision, recall and F1 scores of $0.972$, $0.979$ and $0.975$, respectively. Then, we use this classifier to classify all remaining non-annotated tweets. Finally, we have 141,766 classified misinformation tweets. We crawl all their direct replies, resulting in $793,828$ replies. 
    
Next, we filter tweets to retain those within the scope of our misinformation topics (Section~\ref{sec:misinfotopics}), with at least one of the following (non case sensitive) keywords in the tweet textual string: ``bill gates'', ``fertility'', ``pregnancy'', ``pregnant'', ``gene'', ``dna'', ``gene therapy'', and ``microchip'', resulting in $1,655$ tweets with $26,190$ responses.

To create a high-quality dataset, we manually annotate all the classified $1,655$ misinformation tweets by the textual content to remove false positives and only focus on original tweets (no retweets) and English-language content, as is common practice~\cite{Micallef2020, hossain2020detecting}.

Finally, this dataset contains $798$ misinformation tweets and associated $11,970$ responses.

\subsubsection{\textbf{Annotating Counter-Misinformation Replies and Training the Classifier}}

Naturally, not all responses to misinformation tweets counter it. Therefore, to develop a counter-response dataset, we create the following procedure.

\textit{Training a counter-response classifier:}  Since annotating all $11,970$ responses manually is labor-intensive, we leverage existing work by ~\citet{jiang2020modeling} to create a belief versus disbelief classifier in social media responses.
Specifically, following their pipeline, we create the classifier using RoBERTa~\cite{liu2019roberta} and train it on their annotated responses.
Since the topics of the original data and trained classifier in \citet{jiang2020modeling} are different from ours, we annotated additional responses. 
Specifically, two students annotated $500$ randomly-selected responses from the unlabeled $11,970$ responses, resulting in an inter-rater agreement score of $0.7033$  measured by percent agreement. 
This gave $244$ responses expressing belief and $118$ expressing disbelief, while the remaining were neither expressing belief or disbelief. 
We used these annotated responses to fine-tune the disbelief classifier to our data and topic.
Conducting five-fold cross validation, the classification performances of the classifier per precision, recall and F-1 scores were 0.695, 0.687 and 0.691, respectively. 
Finally, we use the fine-tuned classifier to identify all potential disbelief replies among all the $11,970$ responses. 
This resulted in $2,852$ responses classified as disbelief or counter-response. Then, we manually verify all the classified responses through the textual content to remove all false positives. Finally, $754$ true counter-responses are identified, which we use in our work.

\subsubsection{\textbf{Annotating Linguistic Properties of Counter-Responses}}\label{sec:annotated_counter_responses}

Two students annotated 50 counter-responses as per the three desired properties~\cite{chan2017debunking, steffens2019organisations, malhotra2022meaning}: 
\begin{itemize}
    \item Refuting: is the response explicitly rejecting the false claim or the misinformation spreader?
    \item Evidence: does the response contain evidence or supporting words or sentences to back up the counter-response?
    \item Politeness: Is the reply rude, neutral, or polite like having a soft and friendly tone in the expression?
\end{itemize}

The measured inter-rater agreement score by percent agreement is $78\%$. Disagreements were discussed and a final label was given. Next, each annotator annotated the remaining counter-responses to assign final labels to them. 

Finally, this results in $754$ annotated (misinformation tweet, counter-response) pairs from $238$ misinformation tweets. The distribution of the linguistic properties of counter-responses is shown in Table~\ref{tab:data_stat4_exist_counter}.
\begin{table}[!h]
\small
    \centering
    \begin{tabular}{|c|c||c|c||c|c|}
         \hline
         & Politeness & & Evidence? & & Refutes?   \\ \hline
         Polite  & 51 & Yes &  181 & Yes & 588   \\ \hline
         Neutral & 415 & No & 573 &  No & 166 \\ \hline
         Rude  & 288 & &  &  &  \\ \hline
    \end{tabular}
    \caption{Statistics of $754$ social media counter-responses.}
    \vspace{-9mm}
    \label{tab:data_stat4_exist_counter}
\end{table}

As per the statistics, in-the-wild counter-responses are very low quality -- $38.19\%$ responses are rude, $75.99\%$ do not have evidence, and $22.02\%$ do not explicitly refute the misinformation. This indicates they may not be effective. This further reinforces the critical and timely need for our research to develop an effective counter-response generator.

\subsection{Crowdsourced In-lab Counter-Misinformation  Responses}\label{sec:data_in_lab}

The above statistics show that most in-the-wild responses are rude and lack evidence. As a result, it will be challenging to train an effective counter-response text generator model using this data alone. 
Thus, we create an alternate dataset via crowdsourcing. Motivated by similar text generation for healthy and social good online communication~\cite{ qian2019benchmark, sharma2021towards, tekiroglu2020generating},
we recruit users familiar with Twitter to generate counter-misinformation responses that have the desired properties mentioned earlier in Section~\ref{sec:prob_definition}. 

\noindent \textbf{Ethics:} This protocol was approved by Georgia Tech's IRB.

\noindent \textbf{Procedure:} We use the following three-step process:

First, we recruited 20 college undergraduate and graduate students majoring in engineering domains 
in March 2022. During the screening, subjects provided background information including: (1) Highest education level: high-school, bachelors, masters, or doctorate; (2) Fluency in English: basic, intermediate, advanced (fluent or native speaker); (3) Familiarity with the concept of online misinformation on Twitter: not familiar, somewhat familiar, highly familiar; and (4) Witnessed countering misinformation online: yes or no.

Out of these, 17 participants met the criteria of having least high-school education, being fluent in English, highly familiar with online misinformation, and having seen online debunking.

Second, each subject is provided written guidance about writing an effective counter-misinformation response governed by existing literature~\cite{chan2017debunking, chan2017debunking, steffens2019organisations, malhotra2022meaning}. Representative counter-misinformation examples are shown that are manually selected by the authors from the in-the-wild dataset (Section~\ref{sec:annotated_counter_responses}).
Each subject is given up to 50 randomly-selected COVID-19 vaccine misinformation tweets (from the in-the-wild social media dataset) identified in Section~\ref{sec:misinfo_data_in_the_wild}. 
These tweets span all four misinformation topics (Section~\ref{sec:misinfotopics}) to ensure diverse responses from different subjects. 

After filtering out 90 written responses that do not satisfy any desirable properties (Section~\ref{sec:prob_definition}), we finally created a high-quality counter-misinformation response dataset containing $591$ crowd-generated responses. 
A representative example is shown below:

\begin{figure}[htbp]
\parbox{\linewidth}
{
\parbox{0.93\linewidth}{
\fbox{
\parbox{\linewidth}
{

{
\textbf{\textit{Misinformation Post:}} It’s not a vaccine, it’s gene therapy. Gene therapy is an experimental technique. It’s the same technology used in cloning, DNA editing, and stem cell research.}

\textbf{\textit{In-the-wild Counter-response:}} You are born to speak nothing but lies.

\textbf{\textit{Crowdsourced Counter-response:}} Sorry to see you think in this way. It is not correct. The vaccine is not gene therapy. Instead, it uses mRNA to generate spike protein to protect people. Please do not say the misinformation again.
}}
}
}
\captionsetup{font=scriptsize}
\end{figure}

\section{MisinfoCorrect: A Counter-Response Generation Model}
Here we describe our proposed counter-response generation model that leverages the two datasets to generate counter-responses for a given misinformation post. The generated counter-responses should have the desirable properties described in Section~\ref{sec:prob_definition}. 

\begin{figure}[t]
    \centering
    \includegraphics[width=0.48\textwidth]{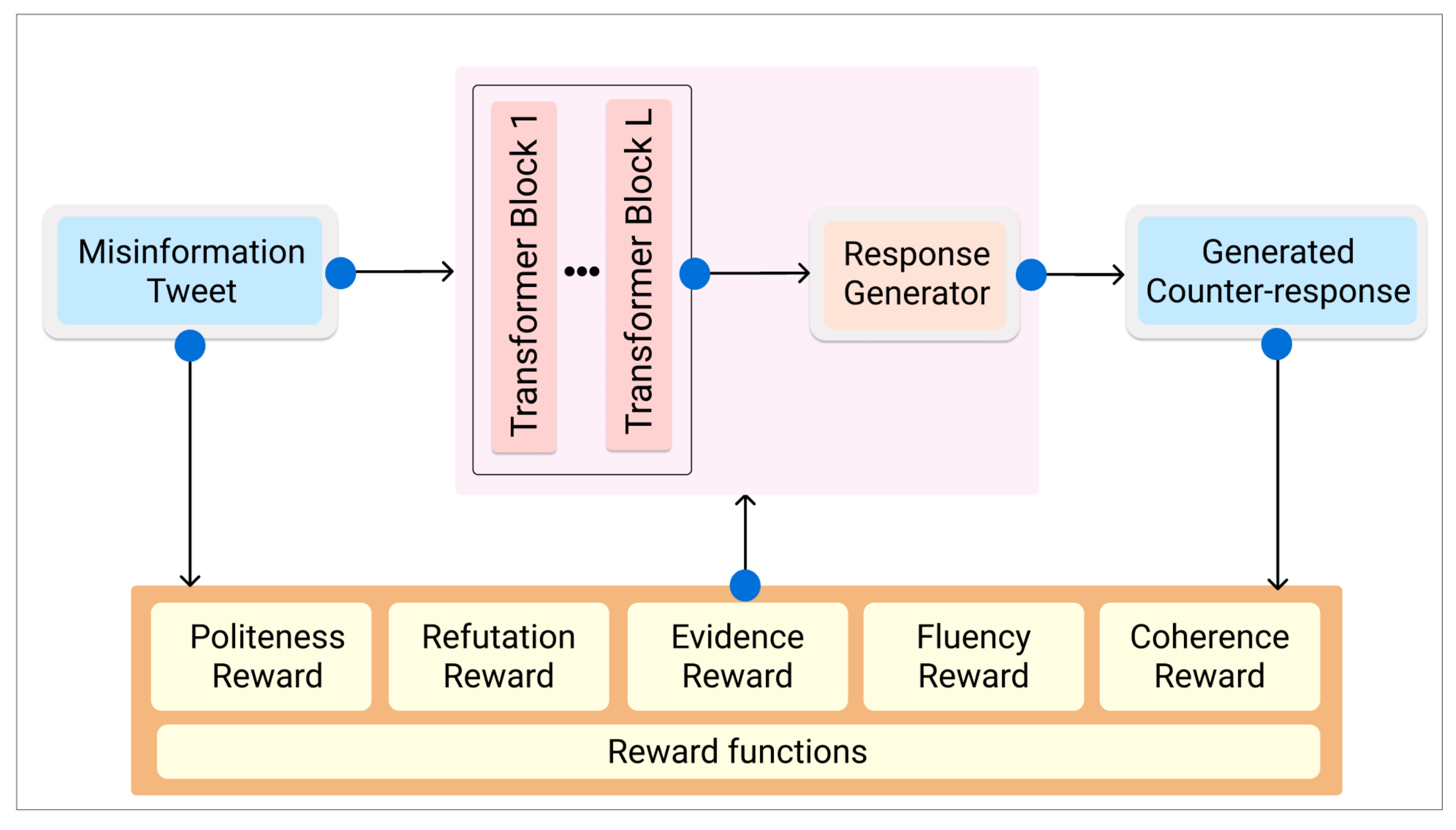}
    \vspace{-7mm}
    \caption{The overview of the MisinfoCorrect framework.  
    }
    \vspace{-5mm}
    \label{fig:model_overview}
\end{figure}

\subsection{A Reinforcement Learning Framework}
We choose a reinforcement learning-based approach due to its success in a variety of controllable text generation tasks~\cite{li2016deep, sharma2021towards}. Moreover, we utilize reinforcement learning (RL) on top of a GPT-2 transformer-based text generation model since it is capable of generating quality example with limited number of  examples derived from its strong generation power and is widely-used in text generation task~\cite{sharma2021towards}. By this design, we can bias the text generation process such that the generated counter-response is of high quality. Figure~\ref{fig:model_overview} presents the overview of MisinfoCorrect. Below we describe the components of the RL agent:

\subsubsection{\textbf{State:}} The misinformation post provides the conversational text. The RL agent takes the misinformation post as the input to enhance the quality of counter-response text so that the response is relevant to the misinformation claims. Formally, the state $s \in \mathcal{S}$ is the same as the content of the misinformation post $m$, i.e., $s = m$. Our policy uses a string containing $s$ for representation, which is also widely used in BERT-like models~\cite{devlin2018bert}.

\subsubsection{\textbf{Action:}}
Given state $s$, the agent generates a candidate counter response $\hat{c}$. This generation action is represented as $a$ lying in the whole action space $\mathcal{A}$, $a \in \mathcal{A}$, which is composed of all arbitrary-length sentences. We represent $g$ as the text generator, and the action is $a = g(s)$.

\subsubsection{\textbf{Policy:}} The policy is based on the transformer language model with the task of masked multi-head self-attention layers on GPT-2~\cite{radford2019language, vaswani2017attention}. The input is an encoded representation of the state $s$ and output is the action $a$. 
The generation task is framed as a language modeling problem where the goal is to generate $\hat{c}$ that maximizes the conditional probability $p(\hat{c}|m)$.
When using transformer component of GPT-2, we first encode our input string ``m''. Then, after transforming the encoded representation as a vocabulary-sized vector using a softmax layer, we have a probability distribution over the entire vocabulary tokens. Next, top-$p$ sampling method is used with the probability distribution to sequentially output a sequence of tokens to form a sentence. When the sampling process selects a special end-of-sequence token, the generation process stops. This generates the candidate counter-response $\hat{c}$.

\subsubsection{\textbf{Reward:}}\label{sec:reward} 
Research has shown that counter-misinformation responses are effective if they are polite, provide evidence, and explicitly refute the misinformation (Section~\ref{sec:prob_definition}).
We design multiple novel reward functions to encourage the generated response to have these properties along with ensuring that the generated text is fluent, coherent, and relevant to the misinformation post. We describe the rewards below.

\noindent $\bullet$ \textbf{Politeness Reward}: Polite counter-responses are preferred (Section~\ref{sec:prob_definition}). 
We quantify the preference toward politeness as a politeness reward ${r}_{politeness}$ and create a politeness classifier $f_{politeness}$ using BERT~\cite{devlin2018bert} to measure politeness of text leveraging existing work~\cite{danescu2013computational}. The classifier fine-tuned and tested in our data in Section~\ref{sec:data} has a classification performance measured via precision, recall and F1 score of $0.8864$, $0.9512$, $0.8001$. The politeness reward is formally computed as ${r}_{politeness} = f_{politeness}(\hat{c})$. 

\noindent $\bullet$ \textbf{Refutation Reward}: Counter-responses that explicitly refute the misinformation are more effective (Section~\ref{sec:prob_definition}). 
Thus, we define the refutation reward ${r}_{refutation}$ to reward the actions that increase refutation of $\hat{c}$ and penalize actions that decrease the refutation of $\hat{c}$. 
Following similar disbelief and polarity classification research works~\cite{agarwal2022graphnli, jiang2020modeling}, we build the refutation classifier $f_{refutation}$ using BERT~\cite{devlin2018bert}  which measures whether the text expresses refutation. 
However, distinct from ~\citet{jiang2020modeling}, who only use the response text for classification, we use both the tweet and generated response as input. The reason is that the refutation relationship would be better predicted by capturing the relative stance between the tweet and its response. We quantify the refutation reward as ${r}_{refutation} = f_{refutation}(m, \hat{c})$.
In our experiments, the refutation classifier is first trained on the annotated data by~\citet{jiang2020modeling}. Then, we fine-tuned and tested it on our data (Section~\ref{sec:data}), which finally achieves reasonable performance in precision, recall and F1 score with values of $0.7917$, $0.8085$, $0.7999$, respectively.

\noindent $\bullet$ \textbf{Evidence Reward}: Responses containing evidence are more effective in countering misinformation~\cite{chan2017debunking}. 
Thus, we seek to generate response that provides textual evidence.
We do not seek to provide a fact-checking URL as evidence, since readers are unlikely to click and read an external article from social media platforms~\cite{glenski2017consumers,glenski2020user}. 
To effectively quantify the presence of evidence in responses, we consider the counter-response content
where the response counters the misinformation post with supporting and relevant sentences. 

We create an evidence classifier $f_{evidence}$ to predict whether the response provides evidence that counters the misinformation post. 
The classifier is trained by combining two sets of evidence-providing responses -- first is the in-the-wild social media counter-responses that contain evidence (Section~\ref{sec:annotated_counter_responses}), and second is the subset of crowdsourced responses (Section~\ref{sec:data_in_lab}) with evidenced responses. Finally, we create a balanced dataset of $573$ evidenced-responses and $573$ non-evidenced-responses to train the classifier. 

We use BERT~\cite{devlin2018bert} as the classifier which takes both the post and response as inputs in a pair-wise setting~\cite{reimers2019sentence} to measure the post-response pairwise relationship.  After five-fold cross validation, the performance score of precision, recall and F1 score is $0.8864$, $0.9512$, $0.9176$. 
The output of the classifier is the evidence reward, ${r}_{evidence}$, computed as ${r}_{evidence} = f_{evidence}(m, \hat{c})$.

\noindent $\bullet$ \textbf{Fluency Reward}: The agent needs to ensure that the response is fluent and grammatically correct. Thus, we want to reward actions that generate fluent outputs and penalize ones that result in non-fluent responses. To achieve this goal, following the previous work~\cite{ma2020powertransformer}, we design the fluency reward ${r}_{fluency}$ which is the inverse of perplexity of the generated countering reply $\hat{c}$ as ${r}_{fluency} = p_{GPT-2}(\hat{c})^{\frac{1}{M}}$, 
where $p_{GPT-2}$ is the GPT-2 language model for English and $M$ is the number of words in $\hat{c}$.

\noindent $\bullet$ 
\textbf{Coherence Reward}: 
Given a misinformation post, the generated response should be relevant to the post. 
We design a coherence reward  ${r}_{coherence}$ which is computed via semantic similarity between $m$ and $\hat{c}$ as ${r}_{coherence} = sim(m, \hat{c})$, where 
$sim$ measures the semantic similarity between two posts. In practice, we utilize the embedding from BERT model of the two text pieces~\cite{devlin2018bert} and compute their cosine similarity.

\noindent
\textbf{Total reward}: Finally, the total reward is as 
\vspace{-3mm}
\begin{multline}\label{equ:attack-loss}
 r = \alpha * r_{politeness} + \beta * r_{refutation} + \gamma * r_{evidence} +  \\
\theta * r_{fluency} +  \lambda * r_{coherence}
\end{multline}
where $\alpha, \beta, \theta, \gamma, \lambda$ are weights indicating the importance of rewards. 

\subsection{Optimization and Training}

\noindent
\textbf{Warm-up start}: We first use the pre-trained weights of DialoGPT \cite{zhang2019dialogpt} to initialize the weights in the transformer-based GPT-2 language model. Next, motivated by the warm-up approaches in reinforcement learning for dialogue generation by ~\citet{li2016deep}, we use the  warm-start strategy on the paired data of misinformation posts and countering replies.

\noindent
\textbf{Reward Increment Training for Reinforcement Learning}: To train the agent in the reinforcement learning framework, we take advantage of the existing reward increment training approach where the non-negative factor, offset reinforcement and characteristic eligibility are considered in the standard reinforcement learning setting~\cite{sutton1999policy}. In our setting for simplicity, we consider the reward $r$ from the generated post and the probability of generating this post given the misinformation post, $p(\hat{c}|m)$. Finally, the loss function $\mathcal{L}$ is computed as $\mathcal{L}(\theta) =  - r * log p(\hat{c} | m)$, 
where $\theta $ is the set of model parameters. We use $log$ to facilitate computation. Meanwhile, we utilize the negative of the reward to deploy the conventional gradient descent approach in experiments. Adam is used as the optimizer for model training~\cite{goodfellow2016deep}.

\section{Experimental Evaluation}
We examine the performance of the proposed counter-misinformation response generation model. In particular, we focus on answering the following research questions:

\noindent $\bullet$ \textbf{RQ1}: Can the proposed model generate counter-misinformation responses of high quality with the desirable properties (Section~\ref{sec:prob_definition})?

\noindent $\bullet$ \textbf{RQ2}: What is the impact of using in-the-wild data versus crowdsourced data on the generated text output? 

\noindent $\bullet$ \textbf{RQ3}: What is the contribution of each component of the proposed method?

\noindent $\bullet$ \textbf{RQ4}: Is the text generated good as evaluated by humans? 

\subsection{Baselines}
We compare our model with representative dialog generation baselines and the work in fact-checking text generation:

\noindent $\bullet$ Fact-checking Text Generation (\textbf{FC-GEN})~\cite{vo2019learning}: The fact-checking text generation model takes in the tweets and replies for generation using gated recurrent unit. 

\noindent $\bullet$ \textbf{DialoGPT}~\cite{zhang2019dialogpt}: A dialogue generation model built on GPT-2 framework and pre-trained on Reddit conversations. 

\noindent $\bullet$ Deep latent sequence model (\textbf{Seq2Seq})~\cite{zhu2018texygen}: A encoder-decoder model for general dialog text generation. 

\noindent $\bullet$ \textbf{BART}~\cite{lewis2019bart}: An large pre-trained language model framework for sequence-to-sequence text generation.

\noindent $\bullet$ \textbf{Partner}~\cite{sharma2021towards}: A reinforcement-learning-based text rewriting method to output text.

\subsection{\textbf{Evaluation Metrics}}
To quantitatively evaluate the performance of the model, we use several metrics to measure both the effectiveness of the counter response and the text quality  as follows:

\noindent $\bullet$ \textbf{Politeness}: We use the politeness classifier $f_{politeness}$  to test the level of politeness expressed in generated responses (Section~\ref{sec:reward}).

\noindent $\bullet$ \textbf{Refutation}: We use the trained refutation classifier $f_{refutation}$  to measure refutation score, as defined in Section~\ref{sec:reward}.

\noindent $\bullet$ \textbf{Evidence}: We use trained evidence classifier $f_{evidence}$  (Section~\ref{sec:reward}) to measure how much evidence the reply provides. 

\noindent $\bullet$ \textbf{Perplexity}: Following previous research~\cite{dai2019style, ma2020powertransformer}, we use pretrained GPT-2 language model to quantify perplexity to evaluate the expressed text fluency.

\noindent $\bullet$ \textbf{Relevance}: Following previous research~\cite{xu2018better}, we compute the semantic similarity using  BERT~\cite{devlin2018bert} to  capture the coherence between posts and generated responses.

\subsection{\textbf{RQ1: Evaluation of the Proposed Model}}

We train all the models with counter-responses from both \textit{social media dataset} (Section~\ref{data:annotated_data_in_the_wild_all}) and \textit{crowdsourced counter-responses} (Section~\ref{sec:data_in_lab}). 
Specifically, we create a ``clean'' social media dataset by only selecting counter-responses with at least one dimension among politeness, refutation, and evidence labeled as positive. This is because training with low-quality counter-responses will lead to poor generation results. 
In addition, we use all crowdsourced counter-responses as they are all manually-verified to be polite, refuting, and evidenced. 

The results comparing the generation models are shown in Table~\ref{tab:social_plus_crowds}.
\begin{table}[!tbp]
    \centering
    \begin{tabular}{|c|c|c|c|c|c|}
    \hline
       \textbf{Method} &  \textbf{Polite. $\uparrow$} & \textbf{Refut. $\uparrow$} & \textbf{Evid. $\uparrow$} & \textbf{Perpl. $\downarrow$}  & \textbf{Rele. $\uparrow$} \\ \hline
       DialoGPT      & 0.874   & 0.831 & 0.693 & 10.010 & 0.930 \\ \hline
       Seq2seq       & 0.794   & 0.794 & 0.621 & 13.403 & 0.948 \\ \hline
       BART          & 0.824   & 0.827 & 0.623 & 11.909 & 0.870 \\ \hline 
       Partner        & 0.892   & 0.898 & 0.702 &  9.781 & 0.871 \\ \hline
       FC-GEN        & 0.815   & 0.714 & 0.594 & 14.971 & 0.810 \\ \hline
       \hline
       MisinfoCorrect  & \cellcolor{blue!25} \textbf{0.915}   & \cellcolor{blue!25}\textbf{0.931} & \cellcolor{blue!25}\textbf{0.723} & \cellcolor{blue!25} \textbf{8.010}  & \cellcolor{blue!25} \textbf{0.960} \\ \hline 
       
    \end{tabular}
    \caption{Performance comparison of counter-response generators when trained on social media and crowdsourced responses. 
    }
    \vspace{-8mm}
    \label{tab:social_plus_crowds}
\end{table}
As can be seen, our proposed model generates the best counter-responses. 
When compared with baselines, our model has the highest politeness, refutation and evidence scores while still maintaining significantly lower perplexity and comparable relevance scores to ensure text of high quality.
Table~\ref{tab:misinfo_running_example} illustrates responses generated by the proposed model and other baselines. As we can see, compared to other methods, MisinfoCorrect can generate text of desirable properties.
\begin{table}[tbp]
    \centering
    \small
    \begin{tabular}{|p{7.8cm}|}
            \hline
         \textbf{Misinformation Post}: It’s not a vaccine, it’s gene therapy. Gene therapy is an experimental technique. It’s the same technology used in cloning, DNA editing, and stem cell research.  \\ \hline
         \textbf{MisinfoCorrect (proposed)}: This is not true. And, the vaccine is not gene therapy. It has nothing to do with cloning or DNA, and only uses mRNA for immunization goal. Please stop this misinformation.  \\ \hline
         \textbf{DialoGPT}: This is so unbelievably wrong. It is not gene therapy. The vaccine does not change DNA. \\ \hline
         \textbf{FC-GEN}: It is misinformation. The vaccine is not gene therapy not gene therapy. \\ \hline
         
    \end{tabular}
    \caption{Examples of generated counter-responses by the proposed and baseline methods.}
    \vspace{-5mm}
    \label{tab:misinfo_running_example}
\end{table}

\begin{table}[tbp]
    \centering
    \begin{tabular}{|c|c|c|c|c|c|}
    \hline
       \textbf{Method} &  \textbf{Polite. $\uparrow$} & \textbf{Refut. $\uparrow$} & \textbf{Evid. $\uparrow$} & \textbf{Perpl. $\downarrow$}  & \textbf{Rele. $\uparrow$} \\ \hline
       DialoGPT      & 0.762  & 0.726 & 0.623 & 12.039 & \cellcolor{blue!25} \textbf{0.940} \\ \hline
       Seq2Seq       & 0.734  & 0.641 & 0.473 & 14.312 & 0.820  \\ \hline
       BART          & 0.723  & 0.721 & 0.607 & 13.079 & 0.893 \\ \hline
       Partner        & 0.781  & 0.709 & 0.632 & 11.993 & 0.825 \\ \hline
       FC-GEN        & 0.714  & 0.663 & 0.515 & 15.102 & 0.782 \\ \hline
       \hline
       MisinfoCorrect     & \cellcolor{blue!25} \textbf{0.854}  & \cellcolor{blue!25} \textbf{0.797} & \cellcolor{blue!25} \textbf{0.643} & \cellcolor{blue!25} \textbf{10.110} & 0.938 \\ \hline
    \end{tabular}
    \caption{Performance comparison of counter-response generators when trained on social media responses only.
    }
    \vspace{-8mm}
    \label{tab:social_text_only_res}
\end{table}

\subsection{\textbf{RQ2: Impact of Dataset Quality}}
Here we examine the impact of the dataset quality on the quality of generated response. 
We train the model using \textit{only a ``clean'' social} media responses (i.e., responses that are evidenced, refuting, neutral, or polite) and no crowdsourced counter-responses. 
The performance results are shown in Table~\ref{tab:social_text_only_res}. 
First, we observe that compared to Table~\ref{tab:social_plus_crowds}, the quality of responses generated by each model degrades. This highlights the importance of collecting crowdsourced data, which is of higher quality compared to social media data. 
Second, we note that our proposed model still generates the best counter-responses as per all metrics, except in relevance, in which it performs the second best.

\subsection{RQ3: Ablation Study}
We examine the contribution of key components for effective counter-response generation (i.e., politeness, refutation and evidence rewards) in MisinfoCorrect on social media and crowdsourced responses data. 
We compare the model variations when using RL:

\noindent $\bullet$ \textit{Base MisinfoCorrect model (Base):} this model is the basic GPT-2 model fine-tuned on our dataset in a dialog manner as DialoGPT~\cite{zhang2019dialogpt}, but without using any rewards for training.

\noindent $\bullet$ \textit{Base + politeness reward:} we only consider the politeness reward

\noindent $\bullet$ \textit{Base + refutation reward:} we only consider the refutation reward

\noindent $\bullet$ \textit{Base + evidence reward:} we only consider the evidence reward.

\noindent $\bullet$ \textit{MisinfoCorrect model:} this is the complete model with all the reward functions. 

The results are shown in Table ~\ref{tab:ablation}.
When we only use the politeness, refutation or evidence reward function in the reinforcement learning framework, the corresponding politeness, refutation and evidence score is the highest and shows a significant increase compared to the Base model without any reward. 
When all the reward functions are combined in the MisinfoCorrect framework, there is a slight drop in each of the individual politeness, refutation, and evidence metrics, but it still has the second highest values along each dimension. This indicates that the MisinfoCorrect model finds a balance between the competing rewards during training.

\begin{table}[tbp]
    \centering
    \begin{tabular}{|p{1.8cm}|c|c|c|c|c|}
    \hline
       \textbf{Method} &  \textbf{Polite. $\uparrow$} & \textbf{Refut. $\uparrow$} & \textbf{Evid. $\uparrow$} & \textbf{Perpl. $\downarrow$}  & \textbf{Rele. $\uparrow$} \\ \hline
       Base & 0.874   & 0.831 & 0.693 & 10.010 & \cellcolor{blue!10}0.930 \\ 
       \text{\hspace{0.1em}}+ politeness   & \cellcolor{blue!25} \textbf{0.953}  & 0.724 & 0.627 &  8.952 &  0.877 \\ 
       \text{\hspace{0.1em}}+ refutation &  0.794  & \cellcolor{blue!25}\textbf{0.968} & 0.623 &  9.138 &  0.856 \\ 
       \text{\hspace{0.1em}}+ evidence &  0.853  & 0.825 &\cellcolor{blue!25}\textbf{0.753} &  \cellcolor{blue!10}8.912 &  0.913 \\ \hline
       MisinfoCorrect  &  \cellcolor{blue!10}0.914  & \cellcolor{blue!10}0.930 & \cellcolor{blue!10}0.723 &  \cellcolor{blue!25}\textbf{8.010} &  \cellcolor{blue!25}\textbf{0.960} \\ \hline 
    \end{tabular}
    \caption{Ablation study.}
    \vspace{-10mm}
    \label{tab:ablation}
\end{table}

\subsection{RQ4: Qualitative Evaluation}

\noindent
\textbf{Experimental Setup:} 
In addition to the quantitative evaluation of response generation, we follow previous research works~\cite{he2021petgen} and  also conducted human evaluation experiments to qualitatively examine the model performance. In particular, we recruited 10 subjects following the same procedure described in the counter-response annotation process (Section~\ref{sec:data_in_lab}). Each subject is presented 30 data points, where each data point consists of one misinformation post and two counter-responses, and then asked ``which response is better when countering the misinformation post: the first, the second, or are they equally effective?''. We test three settings: (1) the real counter-response versus the generated response by MisinfoCorrect; 
(2) the generated response by MisinfoCorrect versus the closest method, i.e., fact-checking generator (FC-GEN)~\cite{vo2019learning}; 
(3) the generated response by MisinfoCorrect versus the most methodologically comparable baseline, i.e., DialoGPT~\cite{zhang2019dialogpt}. We do not inform the subjects which response is generated by which method. Within each setting, we randomly pick 50 data points for comparison, and each data point is annotated by two users. In the analysis of the results, we only summarize the data points on which the two users provide the same label, i.e., disagreement cases are discarded. In total, we received 300 data points in human evaluation  for the three settings.

\noindent \textbf{Ethics:} This protocol was approved by Georgia Tech's IRB.

\noindent 
\textbf{Results}: We get the following result:

(1) \textit{\textbf{Real response versus MisinfoCorrect:}} In 46 out of 50 cases, both annotators provided the same answers. Among these, response generated by MisinfoCorrect were preferred in 76\% cases, while in 6.5\% cases, both responses were rated as equal. Real responses were preferred in the remaining cases.

(2) \textit{\textbf{FC-GEN versus MisinfoCorrect:}} Annotators agreed in 44 out of 50 cases. Among these, MisinfoCorrect was preferred in 61.36\% cases, 18.2\% cases were equal, while 20.5\% responses by FC-generator were better.

(3) \textit{\textbf{DialoGPT versus MisinfoCorrect:}} Annotators agreed in 41 out of 50 cases. Among these, 36.6\% cases prefer MisinfoCorrect, 36.6\% cases are equal, and 26.8\% cases prefer DialoGPT.

From all three comparison results, we can see that responses generated by MisinfoCorrect are preferred over the responses generated by the competing methods and the real responses. One representative example in Table~\ref{tab:misinfo_running_example} also illustrates the difference between these models and real responses.
Altogether, the qualitative results show the potential for MisinfoCorrect in a real application to empower users to counter misinformation.

\section{Discussion and Limitations}

\textit{Generalization across topics, languages, and entire conversations}:
While MisinfoCorrect only studied one topic (COVID-19 vaccine misinformation) on one platform (Twitter) and in one language (English), the proposed model is general. It can be adopted for other topics easily by providing topic-specific data and content from other platforms. Non-English or multi-lingual language models can be used to develop response generator beyond English. Additionally, our method only generates one direct response and does not generate entire conversation (which can be the future work).

\textit{Intended use of the model:} The model can be made available via a web portal or an API, where a user can input a misinformation post and our model will generate one or more counter-responses.

\textit{Backfire effect}: While the backfire effect, i.e., potential increase in misperception due to observing the correction, has been debated for a long time~\cite{lewandowsky2012misinformation, nyhan2010corrections,nyhan2014effective}, many large follow-up studies have failed to replicate backfire effect~\cite{swire2020searching,guess2020does,wood2019elusive}. corrections already has proved to be effective by existing research works~\cite{chan2017debunking,walter2021evaluating,walter2020fact,walter2018unring,porter2021global, seo2022if, friggeri2014rumor}.

\textit{Counter-response may lead to online arguments:}
One may wonder whether using the generated counter-responses can lead to online arguments. 
Our model is intended to encourage users who voluntarily and proactively already counter misinformation to do so in a polite and respectful manner -- recall that 96\% of all counter-misinformation responses are already generated by ordinary users, even though 2 out of 3 times their responses are rude and abusive. 
Since our model generates polite responses, it has lower chance of leading to online fights. 

\textit{Limitation of evaluation based on machine evaluation:} The evaluation replying on classifiers can have limits and are faulty. This may lead to the inaccurate comparison results between models. More human evaluations are needed for a comprehensive comparison.

\section{Conclusion}
Overall, this work shows the potential to build on the recent advancements in generative text models to use them for social good applications. In this work, we extended these models for counter-misinformation response generation. Our proposed model showed promise by generating responses that were qualitatively and quantitatively better than real responses and other generated responses. 

The future work lies in three directions: (i) deploying and evaluating the model in practice, (ii) collecting data from professional fact-checkers as expert-generated counter-responses and compare the model performance against the current setup, and (iii) developing multi-lingual and multi-modal model to generate visual counter-responses.

\noindent \textbf{{ACKNOWLEDGMENTS}}
This research/material is based upon work supported in part by 
NSF grants CNS-2154118, IIS-2027689, ITE-2137724, ITE-2230692, CNS-2239879, and funding from Microsoft, Google, and Adobe Inc. Any opinions, findings and conclusions or recommendations expressed in this material are those of the author(s) and do not necessarily reflect the position or policy of NSF and no official endorsement should be inferred. We thank the CLAWS research group members for their help on the project.


\bibliographystyle{ACM-Reference-Format}
\balance
\bibliography{main}

\appendix
\section{Appendix}

\subsection{Data Annotation and Collection}

The brief guide to writing counter-misinformation responses is:

\begin{figure}[htbp]
\parbox{\linewidth}
{
\parbox{0.93\linewidth}{
\fbox{
\parbox{\linewidth}
{

{
\textbf{\textit{Application Setting:}} 

On Twitter, when someone writes a misinformation tweet, we would like to write a reply to counter the misinformation such that we can mitigate the spread of misinformation.}

\textbf{\textit{Guidance:}} 

Please write a response like you would try to engage or counter the misinformation. When writing replies, you may want to consider the following: \newline
i. You may want to refute the tweet or express disagreement towards the tweet; \newline
ii. You may want to include supporting sentences, reasons, or evidence to make the reply reliable; \newline
iii. You may want to be polite, avoid confrontation, or avoid any impolite or rude expressions in the response.

\textbf{\textit{One Example:}} 

Tweet: ``The Biden "vaccine passport" is here. You either get a non-FDA approved experimental gene modification therapy (euphemistically called a "vaccine") or you'll be denied access to public transportation, sports venues, air travel and more. Obey or stay home.''; 
\newline
Reply: ``To correct you: it is not a gene modification therapy, there is no proof of this nor a scientific rationale, mRNA does not integrate in the human genome. I am for freedom of choice and against vaccine passports but let’s stick to the facts''

\textbf{\textit{One Tweet You Write A Response To:}} 

Tweet: ``It’s not a vaccine, it’s gene therapy. Gene therapy is an experimental technique. It’s the same technology used in cloning, DNA editing, and stem cell research.''
\newline
Your Response: "Sorry to see you think in this way. It is not correct. The vaccine is not gene therapy. Instead, it uses mRNA to generate spike protein to protect people. Please do not say the misinformation again."

}

}
}
}

\captionsetup{font=scriptsize}
\end{figure}

\subsection{Experiment Details}

Some experiment details are included here:
\begin{itemize}
    \item During our experiments, all methods are fine-tuned or trained from scratch on the annotated tweet-response pairs from the in-the-wild and crowdsourced datasets from Section~\ref{sec:data}. 
    \item We cut off responses beyond 280 characters per Twitter’s rule. Note that from Feb 2023, Twitter extends the character limit to 4,000. But, when we did this project, the 280-character limit still held.
    \item For training, MisinfoCorrect takes $7\sim10$ minutes while compared methods use $3\sim10$ minutes. The inference time is comparable across all methods using $0.3\sim0.5$ seconds per example. The one-time longer training ensures higher-quality of generated text. 
    \item In experiments, we set the batch size, training steps, learning rate at 8, 10,000 and 1e-5, respectively. For the the hyperparameter $\alpha, \beta, \theta, \gamma, \lambda$, we used a grid-search-based method with three values (0.1, 1, 10), and we selected $\alpha=1, \beta=1, \theta=10, \gamma=1, \lambda=0.1$. When comparing different methods and reporting results, we take the consistent sampling across all methods for evaluation. 

\end{itemize}

\subsection{Limitations}
\begin{itemize}
    \item \textit{False positives or negatives in misinformation detection classifier may lead to misinformation spread}: We will clearly inform users that the classifier may make mistakes and also provide them links to professional fact checking websites so they can check the content themselves. 

    \item \textit{Bias and partisanship in counter-response behavior by users}: Users tend to follow partisan lines in crowdsourced fact-checking. While the model is not designed to reduce partisan-bias among users, we expect that the model will generate counter-responses only for misinformation posts, regardless of the partisan leaning. 
    
   \item  \textit{Erroneous output and potential for our model's misuse}: 
    To prevent generating unreasonable responses for unknown topics of misinformation or for non-misinformation tweets, we can create a filtering step so that the model will only output a counter-response if the tweet is a misinformation post on the topic(s) it is trained on. 
    
    \item \textit{Harms due to exposure to misinformation in the annotation process}:
    We did not expose ordinary social media users to misinformation.
    Misinformation was shown to crowdworkers (college students) to write counter-responses. 
    We informed the crowd-workers up-front that the content is verified misinformation and they are supposed to write counter-responses. We also provided them with fact-checking resources. This protocol was approved by Georgia Tech's IRB.
    
\end{itemize}

\end{document}